\documentclass{article}
\usepackage{graphicx}
\begin{document}

\title{Shear-Flow Transition: the Basin Boundary}
\author{Norman R. Lebovitz\\Department of Mathematics\\The University
of Chicago\\5734 S. University Ave., Chicago, IL 60637, USA\\email: norman@math.uchicago.edu} \date{} \maketitle

\begin{abstract}\noindent The basin of attraction of a stable
equilibrium point is investigated for a dynamical system (W97) that
has been used to model transition to turbulence in shear flows. The
basin boundary contains a linearly unstable equilibrium point $X_{lb}$
which, in the self-sustaining scenario, plays a role in
mediating the transition in that transition orbits cluster
around its unstable manifold. We find for W97, however, that this role
is played not by $X_{lb}$ but rather by a periodic orbit also lying on
the basin boundary. Moreover, it appears via numerical computations
that all orbits beginning near $X_{lb}$ relaminarize. We offer evidence that this is due to the
exquisite narrowness of the complementary region to the basin of attraction in the part of phase
space near $X_{lb}$. This further leads to a proposal for
understanding the 'edge of chaos' in terms of more familiar invariant
sets of a dynamical system.

\noindent MSC numbers:76D05, 76F20
\end{abstract}

\section{Introduction}\label{intro}
Experimentally, laminar shear flows undergo transition to turbulence
when the relevant parameter, the Reynolds number $R$, exceeds a
critical value $R_c$. Mathematically, when the Navier-Stokes equations
are linearized about the laminar flow, the expected passage from
stability to instability at $R_c$ is not found. This is the familiar
conundrum that linear theory fails to predict the critical value $R_c$
(cf., for example, the introductory remarks in \cite{eck99} for a fuller discussion). A resolution of this conundrum
is that the stable, laminar point $O$ possesses a basin of attraction
$B$ whose boundary $\partial B$ passes increasingly close to $O$ with
increasing $R$, so that perturbations that may be small by laboratory
standards are large enough to transgress $\partial B$ for sufficiently
large values of $R$.

This idea of describing the problem of shear flows in the language of
dynamical-systems theory is an attractive one but has limitations
when the model systems are confined to very low dimensions. For
example, the debate whether turbulence is transient or not
(\cite{eck08}) can hardly be joined in this context, where $\partial
B$ presents a clear boundary between transient and permanent
departures from the laminar flow. The present work is nevertheless
confined to models of very low dimensions. The motivations are (1) the
impression that the nature of the basin of attraction is an important
element in the theory, (2) the observation that very little is known about
it and (3) the conviction that it would be a good idea to understand the
basin and its boundary in low-dimensional systems before proceeding to
high-dimensional systems. In this paper we have considered the
four-dimension model W97 (as described in \cite{W97}) and minor
modifications of it.

Some results that may be relevant to higher-dimensional models and to
the Navier-Stokes (NS) equations, discussed in more
detail in \S \ref{disc}, include the structure of the boundary, which
implies that the functional form of a perturbation may be as
important as its size; that the vicinity of $X_{lb}$ may not be a
good place to seek transition; and that the tendency of the
complementary region to the basin of
attraction to extreme narrowness in some parts of phase space may help
to explain the 'edge of chaos'
(\cite{skufca}) in terms of the more familiar invariant sets of
dynamical-systems theory.

The plan of the paper is as follows. We describe the mathematical
setting in \S \ref{setting}. In \S \ref{w97} we present Waleffe's
model together with diagrams of the boundary of the basin of
attraction indicating the periodic orbit that lies on that boundary,
and the relaminarization of orbits starting near $X_{lb}$. In \S
\ref{strategy} we indicate a resolution of this relaminarization in
terms of the folded structure of the basin boundary. The concluding
section, \S \ref{disc}, is devoted to drawing from this resolution
a proposed interpretation of the edge of chaos and to brief remarks on
the results of this paper and their implications for further study.

\section{Mathematical Setting}\label{setting}
 
The Navier-Stokes (NS) equations possess a number of very simple
solutions representing laminar shear flows (plane Couette and
Poiseuille flow, pipe flow, etc.).  When these partial-differential
equations are modeled by a finite-dimensional system, the laminar flow
can modeled by an equilibrium point of that system, which we'll take
to be the origin of coordinates $O$. Almost all such
finite-dimensional systems that have been studied take the form

\begin{equation}\label{basic_ds}
\dot{x}=Ax + b(x) , x \in R^n \end{equation}
satisfying certain conditions:
\begin{enumerate}
\item \label{nn_crit}$A$ is a non-normal, stable matrix.
\item \label{qq_crit}$b(x)$ is quadratic in $x$ and 
$\sum _{j=1}^n x_j b_j(x) =0.$
\end{enumerate} 
This structure can be inferred by a Galerkin projection of the NS
equations onto $n$ basis vectors, while taking mild liberties with the
boundary conditions.

The stability condition on $A$ implies that its eigenvalues lie in the
left half-plane and therefore that the origin $O$ is asymptotically
stable. The basin of attraction $B$ of an asymptotically stable
equilibrium point is the set with the property that any orbit through
a point of $B$ tends to the equilibrium point as $t \to +\infty$.  It
is an open set invariant under the flow, in the sense that any
solution beginning in $B$ remains in $B$ on its maximal interval of
existence $(a,\infty)$\footnote{In some examples $a=-\infty$, though
this is not inevitable.}. It's boundary $\partial B$ (if it has one)
is likewise invariant in the same sense. The latter is typically of
relative measure zero so orbits that lie {\em on} $\partial B$ are
rare but important since they lie just at the transition from the
laminar flow toward something ``more interesting.'' In particular, the
threshold amplitude for transition, which has been a subject of some
interest (\cite{W05},\cite{cc}, \cite{bt96}), is the minimum distance from the origin to the
basin boundary. We denote such a threshold point by $T$.
\begin{figure}[htp]
\centering \includegraphics[width=5in, height=2in]{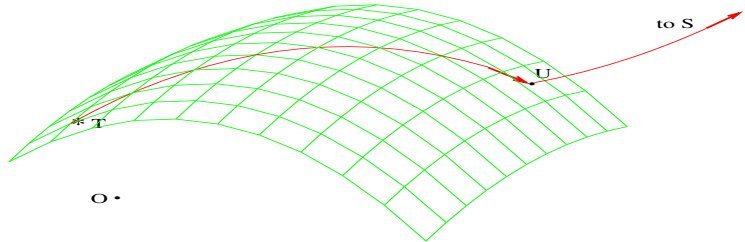}
\caption{\small This cartoon shows Waleffe's picture: a part of the boundary
$\partial B$ of the basin of attraction of the point $O$ is
shown. This part of the boundary coincides with the stable manifold of
the equilibrium point $U$. Orbits starting near $O$ but on $\partial
B$ are attracted toward $U$. Those starting near $O$ but slightly
above $\partial B$ are attracted first toward $U$ but are then
captured by the unstable manifold of the latter and carried away
(toward turbulence). An example is the orbit through the threshold
point $T$.} \label{fw_cartoon}
\end{figure}

\section{The model W97}\label{w97}
This is a four-dimensional model which may be written

\begin{eqnarray}
&& \dot{x}_1=-\delta r_1 x_1 + \sigma _1 x_4^2 - \sigma _2 x_2 x_3,
\label{fws1}\\
&& \dot{x}_2 = - \delta r_2 x_2 + \sigma _2 x_3 + \sigma _2 x_1 x_3 -
\sigma _4 x_4^2, \label{fws2}\\
&& \dot{x}_3 = -\delta r_3 x_3 + \sigma _3 x_4^2 , \label{fws3}\\
&& \dot{x}_4 = -\left(\sigma _1 + \delta r_4\right) x_4 + x_4 \left(
\sigma _4 x_2 - \sigma _1 x_1 - \sigma _3 x_3\right).\label{fws4}
\end{eqnarray}
Here $\delta = 1/R$ where $R$ is the Reynolds number and the eight constants $r_1$ through $\sigma _4$ are all
positive \footnote{In an earlier model,
W95, $\sigma _1$ =0 (\cite{W95}).}. There are standard values for these
constants (cf. \cite{W97}) that we use in this section.

This system conforms to the rules (\ref{nn_crit}) and (\ref{qq_crit})
of model-building. It possesses the symmetry $S=\mbox{diag}(1,1,1,-1)$
so a solution $x(t)$ has a companion solution $\tilde{x}(t)$ obtained
by reversing the sign of $x_4(t)$ so the plane $x_4=0$ is an
invariant plane. For these reasons there is no loss of generality in
considering only solutions for which $x_4(t) \ge 0$. It is not
difficult to show that the invariant plane $x_4=0$ lies entirely in
$B$, the basin of attraction of the origin.

\begin{figure}[htb]
\centering \includegraphics [width=4in, height=2in]{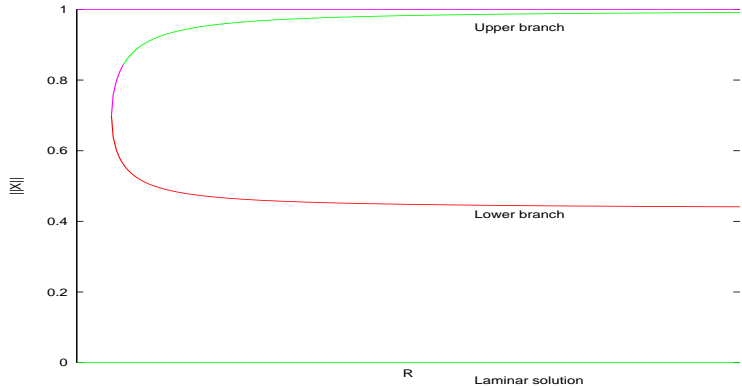}
\caption{\small  
This is the diagram for equilibrium figures in Waleffe's model. The
lower edge, where the norm $\|X\| =0$, represents the laminar
solution. The ``lower branch'' of further equilibrium solutions does
not approach the laminar solution as $R \to \infty$ but instead tends
to the indicated asymptote. However, as shown in \cite{cc}, the curve
of threshold values {\em does} approach the laminar solution, like $R^{-1}$. }\label{fweqfig}
\end{figure}

When one seeks equilibrium solutions other than the origin, they are
found in pairs provided $\delta < \delta _{sn}$ or $R >
R_{sn}$\footnote{The value of $R_{sn}$ of course depends on the other
parameters. With the standard choices for W97, $R_{sn} =
104.84$.}. The one lying closer in norm to the origin is called
the``lower branch'' equilibrium solution $X_{lb}$, the one lying
farther away is called the ``upper branch'' $X_{ub}$ (see Figure
\ref{fweqfig}). The lower-branch solution is
unstable, with a one-dimensional unstable manifold and a three
dimensional stable manifold; the upper-branch solution may be stable
or unstable, depending on the choices of $R$ and of the other
parameters. For the values of $R$ considered in this paper and for the
standard values of the other parameters, $X_{ub}$
is asympotically stable. These equilibrium solutions are illustrated in Figure \ref{fweqfig}.

Figure (\ref{fw_cartoon}) suggests that orbits starting on $\partial
B$ tend toward $X_{lb}$, which mediates the transition. This will be so if the
the stable manifold of $X_{lb}$ coincides with $\partial B$. 
This seems plausible (and has been found to be the case for some other models) but is by no means inevitable: the basin
boundary and the stable manifold of $X_{lb}$ are both invariant sets
for the system (\ref{fws4}) but they are defined in different manners
and need not be identical.
In fact, we find 
that they are not identical for the model W97. The stable manifold of
$X_{lb}$ is a proper subset of $\partial B$, so a point on $\partial B$ lying
far enough from $X_{lb}$ has a different
evolution. 

\begin{figure}[hbt]
\centering \includegraphics [width=5in, height=2in]{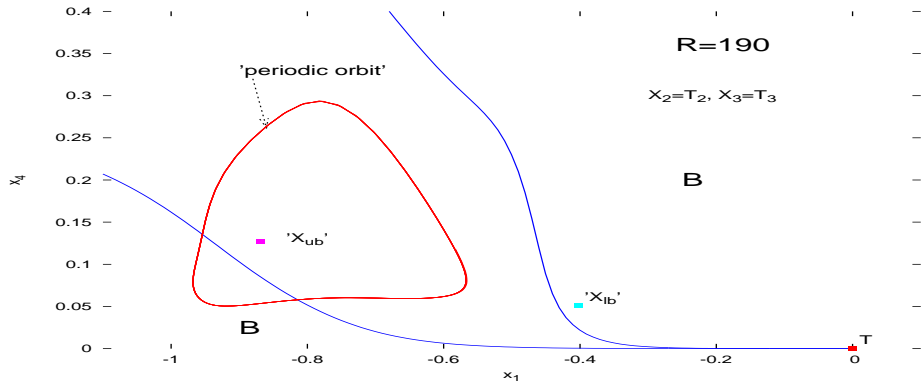}
\caption{\small
A slice through the basin boundary made by the hyperplane $x_2=T_2,x_3=T_3$, where $T_i,
i=1,\ldots,4$ are the components of the threshold point $T$. The parts of
this region lying in $B$ are so marked; the remainder lies outside. An
orbit starting very near $T$ spends a very long time hovering near a
periodic orbit, indicated in projection onto this hyperplane (the
transient leading from $T$ to this orbit is not shown). The
complementary region to the basin
becomes very narrow near $T$ but is easily resolvable numerically for
these values of the parameters. 
.}\label{med190fig}
\end{figure}

\begin{figure}[hbt]
\centering\includegraphics [width=4in, height=2in]{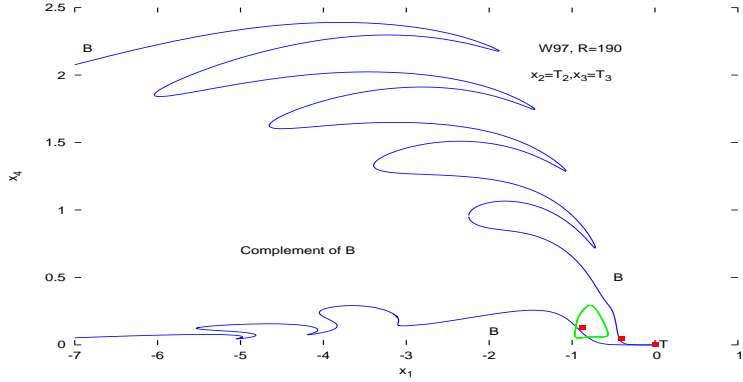}
\caption{\small  
A global view of the preceding figure, showing more of the nature of
the basin boundary, in the same slice as in Figure (\ref{med190fig}).}\label{full190fig}
\end{figure}

\begin{figure}[hbt]
\centering \includegraphics[width=4in, height=2in]{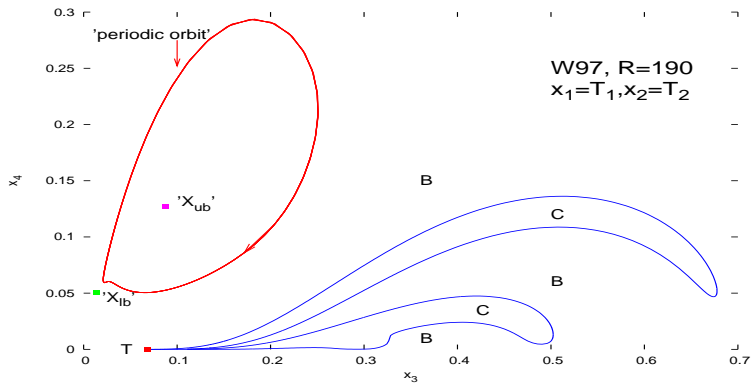}
\caption{\small  
Like Figure (\ref{med190fig}), but a different slice.}\label{full34fig}
\end{figure}

For W97 the threshold point $T$ has been located by Cossu (\cite{cc}). We find,
by following orbits starting near $T$, that they are attracted not
toward $X_{lb}$ but to a periodic orbit $P$ also lying on $\partial
B$.
We have thus far carried out calculations for $R=145$ and $R=190$ and
we display only those for $R=190$ (those for $R=145$ are similar). We
exploit Cossu's calculations to find refinements of the threshold
values $T$: by repeated bisection we obtain a pair of values, $x_o$
and $x_i$ lying respectively just outside and just inside the basin of
attraction $B$ and within a short distance $\epsilon$ of one
another. It follows that there is a point $T$ on $\partial B$ within
$\epsilon$ of either of them. We then find the structure of the basin
boundary by calculating slices of it by various hyperplanes (Figures
\ref{med190fig} and \ref{full34fig}). We also obtain the orbit through
$T$, finding the following.
 On taking $x_o$ as initial data, we find
that after a transient of a few hundred units of time, the orbit is
essentially periodic for many thousands of units of time, eventually
spiraling into the stable, outside point $X_{ub}$; if $x_i$ is taken
as the initial point a similar evolution is found except that at the
end, the orbit tends to the laminar point $O$. Neither of these orbits
comes close to $X_{lb}$.  These remarks are illustrated in
Figures (\ref{med190fig}), (\ref{full190fig}) and (\ref{full34fig}).

We illustrate the results in
the form of slices formed when certain two-dimensional hyperplanes
intersect $\partial B$. Also seen in these diagrams are projections of the periodic
orbit -- and other features -- onto the hyperplanes in question. The
projections are indicated by placing single quotation marks around
their labels.
\medskip

While the self-sustaining process may be only slightly modified by replacing
the equilibrium point $X_{lb}$ with the periodic orbit $P$ as the
mediator of transition, the question of what role
$X_{lb}$ plays in the dynamics now arises. We next turn to this.

\section{Orbits starting near the point $X_{lb}$}\label{strategy}

The
three-dimensional stable manifold of $X_{lb}$ coincides {\em locally} with $\partial B$. Denote by $\xi _t$ a unit vector
transverse to $\partial B$ (for example, $\xi _t$ could have the
unstable direction at $X_{lb}$). Then if for a scalar $v$ we choose initial data 
\begin{equation}\label{unst_pert} x(0) = X_{lb} + v \xi _{t} \end{equation}
for a small value of $|v|$, we expect the orbit to depart from
$X_{lb}$ along its unstable manifold. We anticipate that for one sign
of $v$ the orbit will lie inside $B$ and for the other outside, and
therefore that for one sign the orbit will tend to the origin and for
the other will remain permanently outside the basin boundary, presumably tending
for large $t$ to the stable equilibrium point at $X_{ub}$. 

\begin{figure}[h]
\centering  \includegraphics[width=4in, height=2in]{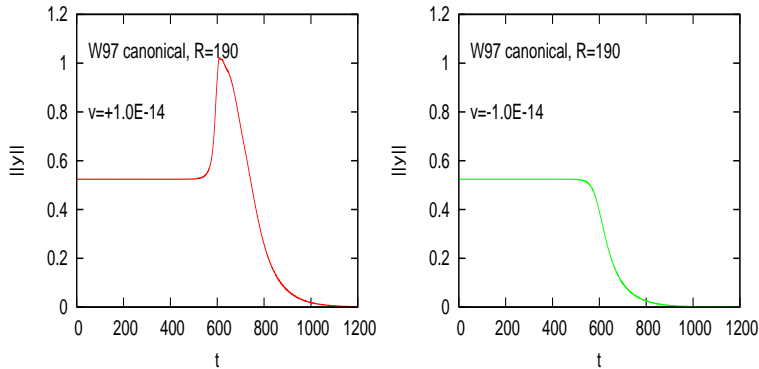}
\caption{\small The model is W97 with the canonical values for the
constants $r_1$ through $\sigma _4$ and $R=190$. The norms of orbits
are shown on the interval [0,1200] and $v= \pm 10^{-14}$. They differ in
the nature of the orbits and in the time
for relaminarization to occur.}\label{norms}
\end{figure}

Instead, we find that all orbits tend to $O$, apparently echoing the persistent
relaminarization found in other models (\cite{eck08}). This is illustrated
in Figure (\ref{norms}).

This violation of expectations could be explained by the following
conjecture for the system W97 near the unstable equilibrium point
$X_{lb}$: $X_{lb}$ indeed lies on $\partial B$ but, near the
part of $\partial B$ on which it lies, there is a second leaf of
$\partial B$
exquisitely close to the first, and it's only for initial data in the
narrow gap between the two leaves that orbits remain bounded away from
the origin \footnote{That narrow gaps are plausible for these
systems may be seen by examining Figure (\ref{med190fig}) near the
point $T$. In that case the gap, while narrow, is still easily detectable numerically.}. If this ``gap conjecture'' is correct for W97 with the
standard choice of parameters, the space between the two leaves is too
small to be detected in the usual double precision arithmetic. The
strategy we adopt below for testing this conjecture is the following.

\begin{figure}[h]
\centering \includegraphics[width=5in, height=2in]{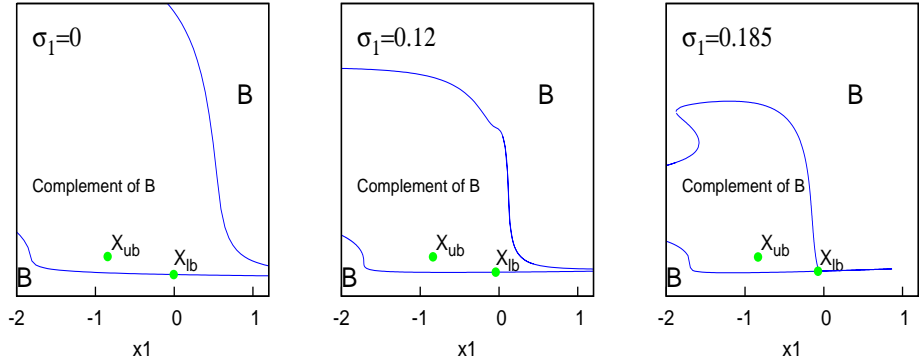}
\caption{\small The ordinate is $x_4$. For each value of $\sigma _1$
shown, there is a progressively narrower gap to the right of
$X_{lb}$. Orbits starting above the narrower gap require longer to
relaminarize than those starting below.
It looks on this scale as if the gap closes when $\sigma _1 = 0.185$, with $X_{lb}$
falling on a critical point. A close up of the region near $X_{lb}$
would show that this has not yet occurred, but see Figure \ref{threecritfig}.}\label{threefig}
\end{figure}

\begin{figure}[hbt]
\centering \includegraphics[width=5in, height=2in]{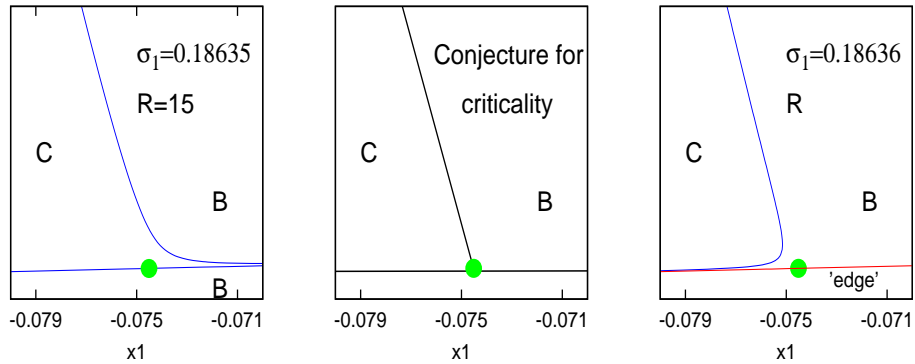}
\caption{\small  The large dots represent $X_{lb}$. These are closeup views of the part of phase space near
$X_{lb}$. The region marked $C$ is the complement of $B$. The first and third of these diagrams are calculated
but the central diagram is conjectural. The curve marked 'edge' in the
third diagram divides long-time relaminarizations from short-time
relaminarizations: no points of the complementary region are
detectable in the region of phase space immediately adjacent to this edge.}\label{threecritfig}
\end{figure}

If we put all the positive constants in W97 equal to unity we find
qualitative similarity to the case with standard values. In
particular, all perturbations of $X_{lb}$ relaminarize. The strategy will
be to take all coefficients equal to unity with the exception of
$\sigma _1$. An asymptotic analysis like that
of \cite{W97} shows that, for small $\delta =1/R$, the lower and upper
branch equilibrium points are
\[ X_{lb} \approx (-\frac{\sigma _1 ^2}{1+\sigma _1 ^2}, \frac{\sigma _1}{1+ \sigma _1
^2},\sigma _1 \delta, \sigma _1 ^{1/2} \delta)\]
and 
\[ X_{ub} \approx (-1 + 2\delta, \delta ^{1/2}, \delta ^{1/2}, \delta
^{3/4}).\]
For small values of $\sigma _1$, we find numerically that there is a large and easily
detectable gap near $X_{lb}$. We then consider successively larger
values of $\sigma _1$ to see if this gap gets successively narrower
and ultimately becomes undetectable. In Figures \ref{threefig} and \ref{threecritfig}, we show only slices of the basin boundary with the
hyperplane $x_2 = X_{lb2}, x_3 = X_{lb3}$ since these seem to reveal
the gap most clearly. The value of $R$ is held fixed at $15$.

These diagrams seem to confirm the gap conjecture but reveal a
further, unexpected feature: there appears to be a topological change
in the nature of the basin boundary for precisely the parameter value
at which the gap becomes suddenly undetectable. We comment further on
this in the discussion section below.

\section{Discussion}\label{disc}

In the self-sustaining scenario for the NS equations, the streaky flow
is represented by a steady-state solution (or by a traveling-wave
solution: steady in a moving frame). These have been sought and
studied in some detail in the context of the NS equations
(\cite{fe03},\cite{wk04}). The analog in W97 is the lower-branch equilibrium
point $X_{lb}$. For standard values of the parameters of W97 we find
that all orbits beginning near $X_{lb}$ relaminarize, whereas in a
different part of phase space -- nearer to the threshold point --
there is a large region wherein orbits are permanently bounded away
from the origin: the low-dimensional analog of persistent
turbulence. Looking for this large region by beginning near $X_{lb}$
would be counter-productive, and it is possible that the analogous
conclusion holds for numerical studies of the NS
equations (cf. the recent exploration of 'relative periodic orbits'
\cite{dpk} in the NS context).

\begin{figure}[hbt]
\centering \includegraphics[width=4in, height=2in]{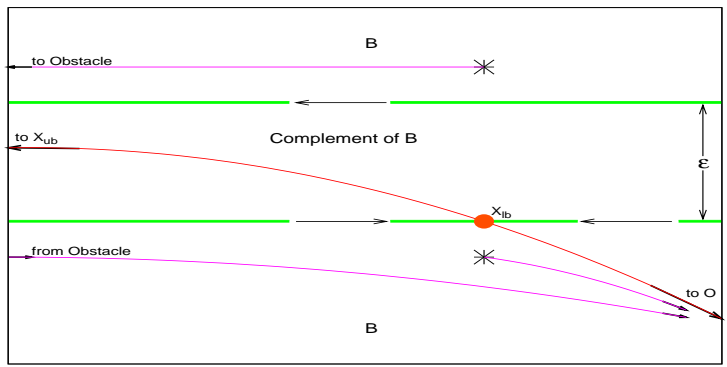}
\caption{\small A conjecture regarding the 'edge.' The distance
$\epsilon$ separating the top and bottom leaves of the basin boundary
is so small as to escape numerical detection. If a point {\em could} be
located between these leaves, it would ultimately be
captured by the stable equilibrium point $X_{ub}$ and therefore be
permanently bounded away from the origin. Attempts to locate
such a point are frustrated by the narrowness of the gap, and will
result instead in initial points indicated by the asterisks above and below. The
orbit starting above is complicated but, since
it is in $B$, ultimately decays to $O$. The orbit starting below
decays more directly, and more quickly, to $O$.}\label{edgefig}
\end{figure}

The apparent collapse of the complementary set to the basin of
attraction, as depicted in Figure (\ref{threecritfig}), is reminiscent
of the 'edge of chaos' as described in (\cite{skufca}) in that the line
marked 'edge' is determined in a similar manner: there is a sharp
difference in relaminarization time for points just above and just
below this line. It suggests a picture like that of Figure
(\ref{edgefig}), wherein the 'edge'\footnote{There is no evidence of
chaos in our calculations so we refer to it simply as the edge.} in
fact consists of two leaves of the basin boundary so close together as
to be numerically indistinguishable.

The diagram shown envisions a region of phase space close to $X_{lb}$
but that is because we have concentrated on this point. There are
undoubtedly parts of
phase space farther from $X_{lb}$ where such narrow gaps occur, 
with behavior that is qualitatively similar but for which the
transient relaminarizations may be quantitatively quite different.

We tentatively propose the structure shown in this diagram as a
building block for the edge of chaos as described elsewhere for other
models (cf. \cite{skufca}). This interpretation would require a
repeated folding and refolding of the basin boundary of exquisite
tightness, each folding modeled by such a building block. That
intricate folds are possible for the basin boundaries of this and
related systems is plausible not only on the basis of the pictures
shown here but also experience with other model systems (e.g., \cite{skufca}). It has been
argued that this 'edge' is itself a new kind of invariant set but it
seems difficult to make this statement mathematically precise. In our
proposed picture, the edge is from any numerical standpoint
effectively invariant since orbits are as close as numerically
possible to the basin boundary, which is indeed invariant.

\bigskip
A number of issues for further study come to mind in view of the
outcomes of the work reported here. We list a few:
\begin{itemize}
\item What is the nature of the singularity indicated in Figure
(\ref{edgefig})? It is possible in principle that the singularity is
an artifact of viewing the surface $\partial B$ through hyperplane
slices and therefore does not represent any singularity of $\partial
B$. Against this is the (apparent) fact that the equilibrium point
$X_{lb}$ of the vector field $f(x)=Ax + b(x)$ lies at the singularity
-- an unlikely coincidence.
\item Can we exploit the picture presented here to locate regions of
phase space where there is {\em no} relaminarization? One could do
that in the context of the present model by following a long-time,
relaminarization orbit, choosing a point on it which is safely far
from both $X_{lb}$ and from $O$, and conducting a search for $\partial
B$ near this point.
\item The structure of the basin boundaries makes it clear that
whether a perturbation lies in $B$ or its complement depends not only
on the amplitude of the perturbation but also -- sensitively -- on its
direction in phase space. Moreover, even for perturbations in the
right direction to leave $B$, a large amplitude is not necessarily
more effective than a smaller one.
\end{itemize}

I wish to thank Carlo Cossu for generously sharing his data with me.

\bibliography{basin}
\end{document}